\documentclass[12pt]{article}
\usepackage[margin=2cm]{geometry}
\usepackage{comment}
\usepackage{amsmath,amssymb,extarrows,mathtools,graphicx,subfigure,setspace}
\usepackage{cite}
\usepackage{slashed}
\usepackage{color}
\makeatother

\numberwithin{equation}{section}

\newcommand{\be}{\begin{equation}}
\newcommand{\bea}{\begin{eqnarray}}
\newcommand{\eea}{\end{eqnarray}}
\newcommand{\ba}{\begin{array}}
\newcommand{\ea}{\end{array}}
\newcommand{\ee}{\end{equation}}

\begin{document}
\onehalfspacing
\noindent
\begin{titlepage}
\hfill
\vspace*{20mm}
\begin{center}
{\Large {\bf Flat-Space Energy-Momentum Tensor  \\ from BMS/GCA Correspondence }\\
}

\vspace*{15mm} \vspace*{1mm} \textbf{{Reza Fareghbal,\;   Ali Naseh }} 

 \vspace*{1cm}
{\it  School of Particles and Accelerators,\\
 Institute for Research in Fundamental Sciences (IPM)\\
P.O. Box 19395-5531, Tehran, Iran \\ }

\vspace*{.4cm}

{E-mails: {\tt fareghbal@theory.ipm.ac.ir, naseh@ipm.ir}}%

\vspace*{2cm}
\end{center}
\begin{abstract}
Flat-space limit is well-defined for asymptotically AdS spacetimes written in coordinates called the BMS gauge. For the three-dimensional Einstein gravity with a negative cosmological constant,  we calculate the quasi-local energy momentum tensor  in the BMS gauge and take  its  flat-space limit. In defining the flat-space limit, we use the BMS/GCA correspondence which is a duality between gravity in flat-spacetime and a field theory with Galilean conformal symmetry.   The resulting  stress tensor reproduces correct values for conserved charges of three dimensional asymptotically flat solutions. We show that the conservation relation of the flat-space energy-momentum tensor is given by an ultra-relativistic contraction of its relativistic counterpart. The conservation equations correspond to Einstein equation for the flat metric written in the BMS gauge. Our results provide further checks for the proposal that the holographic dual of asymptotically flat spacetimes is a field theory with Galilean conformal symmetry.

\end{abstract}
\end{titlepage}
\section{Introduction}

One of the outstanding questions in modern theoretical physics is understanding the holographic principle. For this, it is of interest to explore whether holography exists beyond the known example of AdS/CFT and for spacetimes other than AdS. The study of holography for asymptotically flat spacetimes may be the first step in this direction due to simplicity of flat space and its close relation with the real physical world.  During the past two decades, there have been several attempts at trying to formulate flat space holography, but there have been several hurdles to this apparently simple problem. 

A new method for studying this problem was proposed recently  in \cite{Bagchi:2010zz, Bagchi:2012cy}. This was based on the fact that the  asymptotic symmetry group (ASG)  of asymptotically flat spacetimes at null infinity is the Bondi, van~der Burg, Metzner and Sachs (BMS) group \cite{BMS}. In three and four dimensions, BMS group is infinite dimensional and the corresponding algebra can be centrally extended \cite{bar-comp}-\cite{Barnich:bmscft1}. Thus it is probable that the theory dual to three or four dimensional flat space is given by a (two or three dimensional) field theory living on its null boundary which has exactly the symmetries of the BMS group. 

The observation of \cite{Bagchi:2010zz} was that BMS algebra is isomorphic Galilean conformal algebra (GCA) which is a result of contraction of conformal algebra \cite{GCA}. This connection dubbed the BMS/GCA correspondence was studied carefully in \cite{Bagchi:2012cy} where it was shown that in order to have a well-defined correspondence at the level of centrally extended algebras, at the level of spacetime, the time direction must be contracted in the CFT and the resultant theory is an ultra-relativistic field theory. In other words, flat space limit of AdS which in the bulk side implemented by taking the large radius limit of asymptotically AdS spacetimes is equivalent to the contraction of time on CFT in the boundary side. In this way one can study holography of flat spacetimes just by starting from AdS/CFT and taking appropriate limit: large radius limit in the bulk and contraction in the boundary \cite{Bagchi:2012cy}.
  
This method has met with some very interesting recent successes. It was shown in \cite{Bagchi:2012xr} that a Cardy-like formula for field theories with Galilean conformal symmetry in two dimensions (resulting from time-contraction of two dimensional CFT on the cylinder) produces the same entropy as that of three-dimensional cosmological solutions of flat gravity. These solutions of Einstein gravity in three dimensions were first studied in \cite{Cornalba:2002fi} and are shift-boost orbifolds of three dimensional Minkowski spacetime. They can be viewed as the large radius limit of non-extremal BTZ black holes in AdS$_3$.  Moreover, it was shown in \cite{Bagchi:2013lma} that there is a Hawking-page like transition between Minkowski and flat cosmological solutions. The limiting procedure has been repeated for the topologically massive gravity in three dimensions and the corresponding two dimensional unitary theory has been recognized \cite{Bagchi:2012yk}. A link between GCA and the symmetries of the tensionless limit of closed bosonic string theory has been shown in \cite{Bagchi:2013bga} and finally a similar contraction has been found for the $W_3$  algebra which extends the BMS algebra to include spin-three fields in gravity three dimensional flat spacetimes\cite{Afshar:2013vka}.   
 
A subtle point in using proposal of \cite{Bagchi:2010zz, Bagchi:2012cy} for flat-holography is that the flat space limit in the bulk is not well-defined for all coordinates of asymptotically AdS spacetimes, e.g. the Fefferman-Graham coordinate system. However, \cite{Barnich:2012aw} proposed a new coordinate system for asymptotically AdS spacetimes where taking flat-space limit is well-defined. In these coordinates, called BMS gauge, one can define appropriate fall-off conditions for the three dimensional metric components and find the asymptotic symmetry group in a manner similar to the seminal method of \cite{Brown:1986nw}.
 
In this paper we show that writing asymptotically AdS spacetimes in BMS gauge gives more insight about holography of asymptotically flat spacetimes. A natural question to ask is whether bulk computations in flat space can yield the correlations of the boundary theory with the symmetries of a 2d GCA. The AdS/CFT correspondence provides a beautiful recipe for finding such correlators.  If we follow methods of \cite{Bagchi:2010zz} and \cite{Bagchi:2012cy} for flat-space holography, it is plausible that taking flat limit of bulk calculations in the AdS case gives correlators which corresponding to the contracted correlators in the CFT side. As the first step in this direction we apply  the method of \cite{Balasubramanian:1999re} to compute the one-point function of energy-momentum tensor and by writing it in the BMS gauge we define an appropriate flat limit for the quasi-local energy-momentum tensor.  Our definition of stress tensor for flat holography is consistent with the results of a field theory with Galilean conformal symmetry.   

Our final answer for the one point function of energy-momentum tensor can be used in the calculation of conserved charges of bulk solutions. In three dimensional flat gravity  the most important solutions are cosmological solutions with a cosmological horizon and finite entropy, mass and angular momentum. Our energy- momentum tensor reproduces the correct values for the mass and angular momentum of these solutions. 

This paper is organized as follows. In the next section we start from asymptotically AdS$_3$  spacetimes written in the BMS gauge and introduce its generic solution. The proper flat limit of the BMS gauge is also discussed  in this section.   A brief introduction to BMS/GCA correspondence is given in section 3. The quasi-local energy momentum tensor in the BMS gauge together with a well-defined limit of it are developed in section 4. The derivation of conserved charges of solutions of three dimensional flat gravity and also the relation with one point functions of field theory with Galilean conformal symmetry are also presented in section 4.

\textbf{Note added:} While this paper was being prepared for submission, the work \cite{Costa:2013vza} appeared on arXiv which has some overlap with our work. In \cite{Costa:2013vza}, the author has performed a holographic renormalization of AdS gravity in the BMS gauge and taken its flat limit. We believe that the two pieces of work agree where there is overlap and in our paper the connection of flat limit with field theory with Galilean conformal symmetry is highlighted. This is novel and has not been addressed in \cite{Costa:2013vza}.

\section{AdS$_3$ in BMS gauge and its flat limit}
Instead of Fefferman-Graham gauge which is a common way of defining asymptotically AdS space-times one can write a generic solution of three dimensional Einstein gravity with cosmological constant,
\begin{equation}
S={1\over 16\pi G}\int\, d^3x \sqrt{-g}(R+{2\over\ell^2}),
\end{equation}
 in a new coordinate system as \cite{Barnich:2012aw}
\begin{equation}\label{BMS gauge}
ds^2=e^{2\beta} {V\over r}du^2-2e^{2\beta} du dr +r^2(d\phi-U du)^2,
\end{equation}
where $\beta, V, U$ are arbitrary functions of coordinates and $u$ is a null coordinate known as \textit{retarded time}. This coordinate-system is known as BMS coordinate and it is possible to find the appropriate coordinate transformation for writing all asymptotically AdS solutions in this form.  For AdS$_3$ in the global coordinate 
\begin{equation}
ds^2=-(1+{r^2\over\ell^2})dt^2+{dr^2\over 1+{r^2\over\ell^2} } +r^2 d\phi^2,
\end{equation}
$u$ is given by $u=t-\ell\arctan{r\over\ell}$. It is clear that the boundary of AdS$_3$ is at $r=\infty$. For asymptotically AdS$_3$ spacetimes  $\beta=U={o}(1)$. Thus one can solve the equations of motions and find 
\begin{equation}\label{AdS in BMS}
ds^2=\left(-{r^2\over\ell^2}+\mathcal{M}\right)du^2-2dudr+2\mathcal{N}dud\phi+r^2 d\phi^2,
\end{equation}
where
\begin{equation}\label{MN AdS}
\mathcal{M}(u,\phi)=2\left(\Xi(x^+)+\bar\Xi(x^-)\right),\qquad \mathcal{N}(u,\phi)=\ell\left(\Xi(x^+)-\bar\Xi(x^-)\right),
\end{equation}
and $\Xi,\bar\Xi$ are arbitrary functions of $x^\pm={u\over\ell}\pm\phi$.

The  Killing vectors which preserve the form of line element \eqref{AdS in BMS} are 
\begin{equation}\label{AKV AdS}
\xi^u=f,\qquad \xi^\phi=y-{1\over r}\partial_\phi f, \qquad \xi^r=-r\partial_\phi y+\partial_\phi^2f-{1\over r} \mathcal{N}\partial_\phi f,
\end{equation}
where
\begin{equation}
f={\ell\over 2}(Y^+(x^+)+Y^-(x^-)),\qquad y={1\over 2}(Y^+(x^+)-Y^-(x^-)),
\end{equation}
$Y^+$ and $Y^-$ are arbitrary functions of $x^+$ and $x^-$. At the leading term the generators  
\begin{equation}
{\mathcal{L}}_n=\xi\left(Y^+=-i\exp(in x^+),Y^-=0\right),\qquad \bar{\mathcal{L}}_n=\xi\left(Y^+=0,Y^-=-i\exp(in x^-)\right),
\end{equation}
satisfy  the algebra 
\begin{equation}\label{Virasoro}
[\mathcal{L}_m,\mathcal{L}_n]=(m-n)\mathcal{L}_{m+n},\qquad [ \mathcal{\bar L}_m,\mathcal{\bar L}_n]=(m-n)\mathcal{\bar L}_{m+n},\qquad [\mathcal{L}_m,\mathcal{\bar L}_n]=0.
\end{equation}
In the level of  conserved charges, the above algebra has a central extension with central charges $c=\bar c= 3\ell/2G$ \cite{Barnich:2012aw}.

The flat limit of asymptotically AdS spacetimes written in the BMS gauge is well-defined. This is done by taking $\ell\to\infty$ limit of the AdS case. Precisely, we will take $G/\ell\to 0$ while keeping $G$ fixed. The starting point is the metric \eqref{AdS in BMS} and generic forms for functions $\mathcal{M}$ and $\mathcal{N}$ in \eqref{MN AdS}. The Fourier expansion of  functions $\Xi$ and $\bar\Xi$ are
\begin{equation}
\Xi(x^+)=\sum_{n=-\infty}^\infty f_n\left({G\over\ell}\right)e^{inx^+},\qquad \bar\Xi(x^-)=\sum_{n=-\infty}^\infty \bar f_n\left({G\over\ell}\right)e^{inx^-}.
\end{equation}
We assume that functions $f_n$ and $\bar f_n$ have well-defined behaviour at $\ell\to\infty $. Thus one can write an expansion for them in terms of  positive powers of $G/\ell$. If we demand that 
\begin{equation}
M=\lim_{{G\over\ell}\to 0} \mathcal{M}, \qquad N=\lim_{{G\over\ell}\to 0} \mathcal{N},
\end{equation}
are well-defined, the coefficients  of $(G/\ell)^0$ in $f_n$ and $\bar f_{-n}$ must be the same. Then, it is not difficult to check that 
\begin{equation}\label{MN flat}
M=\theta(\phi),\qquad N=\chi(\phi)+{u\over 2}\theta '(\phi),
\end{equation}
where $\theta$ and $\chi$ are arbitrary functions. Thus the asymptotically flat solution is
\begin{equation}\label{flat  in BMS}
ds^2= M du^2-2dudr+2{N}dud\phi+r^2 d\phi^2.
\end{equation}
The line element \eqref{flat  in BMS} with condition \eqref{MN flat} is exactly  the generic result which one may find for solutions of Einstein gravity (without cosmological constant) in the BMS gauge \eqref{BMS gauge}\cite{Barnich:bmscft1}. We should emphasize that we don't use the modified Penrose limit of \cite{Barnich:2012aw} which needs the scaling of the Newton constant $G$ and results in an ambiguity in the definition of conserved charges\footnote{ A new   Grassmannian method    for mapping gravity in AdS$_3$ to  gravity in flat spacetime was introduced recently in \cite{123}.    }.  

Similarly, the  Killing vectors of flat case can be given by taking $G/\ell\to 0$  from \eqref{AKV AdS} by using  Fourier expansions of functions $Y^+$ and $Y^-$. The final result is 
\begin{equation}\label{AKV flat}
\xi^u=F,\qquad \xi^\phi=Y-{1\over r}\partial_\phi F, \qquad \xi^r=-r\partial_\phi Y+\partial_\phi^2F-{1\over r} {N}\partial_\phi F,
\end{equation}
where
\begin{equation}
Y=\lim_{{G\over\ell}\to 0} y=Y(\phi),\qquad F=\lim_{{G\over\ell}\to 0} f=T(\phi)+u Y'(\phi).
\end{equation}
$Y$ and $T$ are arbitrary functions.  This result again coincides with direct calculations of \cite{Barnich:bmscft1}. 

Using \eqref{AKV flat}, one can find asymptotic symmetry group of flat spacetimes. The important point which we should emphasize here is that the u-coordinate of BMS gauge for flat and AdS spacetimes characterizes different hyperspaces from the causal structure  point of view. As  mentioned earlier, $(u,\phi)$ for AdS$_3$ determine spatial boundary of spacetime, however $(u,\phi)$ for the flat case are the coordinates of future null infinity $\mathcal{I}^+$. This fact  is easily seen by the definition of $u=t-\ell\arctan{r\over\ell}$ in the AdS case which is transformed to $u=t-r$ by taking the $\ell\to\infty$ limit.  Defining generators as 
\begin{equation}
L_n=\xi\left(Y=-i\exp(in \phi),T=0\right),\qquad M_n=\xi\left(Y=0,T=-i\exp(in \phi)\right),
\end{equation}
one can easily check that in the leading term we have 
\begin{equation}\label{BMS3}
[L_m,L_n]=(m-n){L}_{m+n},\qquad [ L_m,M_n]=(m-n)M_{m+n},\qquad [M_m,M_n]=0.
\end{equation}
This algebra is known as BMS$_3$ algebra \cite{Ashtekar:1996cd}. The interesting point is that the algebras \eqref{Virasoro} and \eqref{BMS3} are related by 
\begin{equation}\label{relation virasoro BMS}
L_n=\mathcal{L}_n-\mathcal{\bar L}_{-n},\qquad M_n={G\over\ell}\left(\mathcal{L}_n+\mathcal{\bar L}_{-n}\right),
\end{equation}
in the limit $G/\ell\to 0$. In fact the relation \eqref{relation virasoro BMS} is also applicable for \eqref{Virasoro} with central extension. If one uses Brown and Henneaux's central charges $c=\bar c= 3\ell/2 G$, then the algebra \eqref{BMS3} has a non-zero central extension in the $[L_m,M_n]$ part.   

\section{BMS/GCA correspondence}\label{BMS/GCA}

Interestingly, the algebra \eqref{BMS3} also appears in another branch of higher energy physics. If one considers a two dimensional CFT  and takes its non-relativistic limit by contracting one of the coordinates, the contracted algebra is exactly \eqref{BMS3}\cite{GCA}.  In this context   the algebra \eqref{BMS3} is  called Galilean conformal algebra due to appearance of Galilean  subalgebra. Because of this similarity, it was proposed in \cite{Bagchi:2010zz} that the dual theory for the asymptotically flat space times is a non-relativistic conformal filed theory and this duality was coined as BMS/GCA correspondence.

The parent CFT in the study of  \cite{Bagchi:2010zz} is  on the plane  and the contracted coordinate was the $x$-coordinate. This assumptions dictated  some restriction on the centrally extended algebra in such a way that in order to reproduce the known results of flat-space gravity the parent CFT must be non-unitary. This problem was reconsidered in the paper \cite{Bagchi:2012cy} and the correct way of constructing the dual theory of flat-space gravity was proposed.

The main idea of  \cite{Bagchi:2012cy} is that the CFT must be on a cylinder and time coordinate is the appropriate coordinate for contraction. The importance of cylinder versus plane is clear from the fact that in the bulk side the flat limit is only well-defined for the global AdS. Another way of introducing the contracted coordinate is that we should contract the non-compact coordinate of the boundary theory. 

The states for the field theory with Galilean conformal symmetry is given by $|h_L,h_M\rangle$ where $h_L$ and $h_M$ are the labels of the states and are eigenvalues of $L_0$ and $M_0$. There exists a notion of primary states in this theory and the representations are built by acting with raising operators $L_{-n}$ and $M_{-n}$ on these primary states. Similar to usual 2d CFTs, one can find a Cardy-like formula for the entropy of states $|h_L,h_M\rangle$. This calculation was done in \cite{Bagchi:2012xr} and the final entropy for large $h_{L}$ and $h_{M}$ is
\begin{equation}\label{Cardy-like}
  S =  \ln d(h_L, h_M) =  \pi\bigg( C_{LL}
\sqrt{\frac{2h_M}{C_{LM}}} + 2h_L \sqrt{\frac{C_{LM}}{2h_M}}
\bigg).
  \end{equation}  
where $C_{LL}$ and $C_{LM}$ are the central charges which appear respectively in the $[L_m,L_n]$ and $[L_m,M_n]$ parts of the algebra \eqref{BMS3}\footnote{The leading correction to this formula was introduced recently in \cite{Bagchilast}.  }. It was shown in \cite{Bagchi:2012xr} that \eqref{Cardy-like} gives the correct entropy for the cosmological horizon of the three dimensional Einstein gravity solution which is the flat limit of BTZ black holes. This is an evidence in favour of the correctness of BMS/GCA correspondence.

\section{Definition of  energy-momentum tensor for asymptotically flat spacetimes }
The fact that the flat space limit is well-defined in the BMS gauge tempts us to repeat all holographic calculations in these coordinates and take its flat limit. As a first step, we would like to study the method of holographic renormalization and find correlation functions of energy momentum tensor of the dual theory \cite{HH}. The warm up calculation is finding one-point function of energy momentum tensor which can be used in the bulk side for the calculation of conserved charges. The simple method for this calculation is given by \cite{Balasubramanian:1999re} and \cite{HH}  which uses Brown and York's proposal \cite{Brown:1992br} for the definition of quasi-local stress tensor. According to \cite{Brown:1992br}, the Brown and York's quasi-local energy-momentum tensor is given by 
\begin{equation}
 T^{\mu\nu} = \frac{2}{\sqrt{-\gamma}}\frac{\delta S}{\delta\gamma_{\mu\nu}},
 \end{equation} 
where $S= S_{grav}(\gamma_{\mu\nu})$ is the gravitational action viewed as
a functional of boundary metric $\gamma_{\mu\nu}$\footnote{This is also the standard formula
for the stress tensor of field theory with action $S$ that lives on a background with metric
$\gamma_{\mu\nu}$.}. To be more precise, the gravitational action is
\begin{equation}
S = \frac{1}{16\pi G}\int _{\mathcal{M}} d^{3}x \sqrt{-g}
\left( R- \frac{2}{\ell^{2}}\right) -\frac{1}{8\pi G} \int_{\partial\mathcal{M}}
d^{2}x \sqrt{-\gamma} \;\mathcal{K}+\frac{1}{8\pi G}S_{ct} (\gamma_{\mu\nu}),
\end{equation}
where the second term is known as Gibbons-Hawking term and is
required for a well defined variational principle. The $S_{ct}$ is the
counterterm action that we must add in order to obtain a finite stress tensor. Moreover, 
$\mathcal{K}$ is trace of the extrinsic curvature of the boundary
\begin{eqnarray}
\mathcal{K} &=& \gamma^{\mu\nu}\mathcal{K}_{\mu\nu} \;=\;
\gamma^{\mu\nu}\gamma_{\mu} ^{\rho}\;\nabla_{\rho}n_{\nu},
\end{eqnarray}
where $\gamma_{\mu\nu} = g_{\mu\nu}-n_{\mu}n_{\nu}$ is the boundary metric 
and $n_{\nu}$ is the outward pointing normal vector to the boundary $\partial\mathcal{M}$. \\

Using  above definitions, the quasilocal stress tensor becomes
\bea\label{stress.tensor}
T_{\mu\nu} &=& -\frac{1}{8\pi G} \left( \mathcal{K}_{\mu\nu}
-\mathcal{K}\gamma_{\mu\nu} +\frac{2}{\sqrt{-\gamma}}\frac{\delta S_{ct}}
{\delta\gamma^{\mu\nu}}\right),
\eea
where the counterterm action in three dimensional bulk is \cite{Balasubramanian:1999re,HH}
\bea\label{counterterm.action}
S_{ct} &=& -\frac{1}{\ell}\int _{\partial\mathcal{M}}\sqrt{-\gamma}.
\eea

Let us start from \eqref{AdS in BMS} and  by using \eqref{MN AdS} write it in the light-cone coordinates $x^\pm=\frac{u}{l}\pm\phi$,
\begin{equation}\label{BMS in ligh-cone}
\begin{split}
ds^2=& {r^2\over G^2}\left\{-G^2dx^+dx^-+\frac{\ell^2 G^2}{r^2}\left[\Xi (dx^+)^2+\bar\Xi (dx^-)^2+(\Xi+\bar\Xi)dx^+ dx^-\right]\right\}\cr
& -\ell(dx^+ + dx^-)dr.
\end{split}
\end{equation}
Using (\ref{counterterm.action}) we have
\bea
T_{\mu\nu}^{ct} &=& -\frac{1}{8\pi G \ell} \gamma_{\mu\nu},
\eea
therefore (\ref{stress.tensor}) reduces to
\bea\nonumber
T_{rr} &=&\mathcal{O}(\frac{1}{r^{2}}),\;\;\;\;\;\;T_{r+}=\mathcal{O}(\frac{1}{r^{2}}),
\;\;\;\;\;\;T_{r-}=\mathcal{O}(\frac{1}{r^{2}}),\;\;\;\;\;\;
T_{+-} = \mathcal{O}(\frac{1}{r^{2}})\\ 
T_{++} &=& \frac{\ell}{8\pi G}\Xi(x_{+})+\mathcal{O}(\frac{1}{r}),\;\;\;\;\;\;T_{--} =
\frac{\ell}{8\pi G}\bar{\Xi}(x_{-})+\mathcal{O}(\frac{1}{r}).\nonumber \\
\eea
Thus we find the energy-momentum one-point function of dual CFT as
\begin{equation}\label{EMT AdS light cone}
\langle T_{++}\rangle= \dfrac{\ell\Xi}{8\pi G},\qquad \langle T_{--}\rangle= \dfrac{\ell\bar \Xi}{8\pi G},\qquad \langle T_{+-}\rangle= 0.
\end{equation}
In the coordinate $\{u,\phi\}$ we have 
\begin{equation}\label{EMT AdS uphi}
\langle T_{uu}\rangle= \dfrac{\Xi+\bar\Xi}{8\pi G\ell}=\dfrac{\mathcal{M}}{16\pi G\ell},\qquad  \langle T_{u\phi}\rangle= \dfrac{\Xi-\bar\Xi}{8\pi G}=\dfrac{\mathcal{N}}{8\pi G\ell},\qquad \langle T_{\phi\phi}\rangle= \dfrac{\ell(\Xi+\bar\Xi)}{8\pi G}=\dfrac{\ell\mathcal{M}}{16\pi G}.
\end{equation}

Since the $\ell\to\infty $ limit of $\mathcal{M}$ and $\mathcal{N}$ is well-defined, it is obvious from \eqref{EMT AdS uphi} that defining energy-momentum tensor of flat-space holography by taking the limit from the AdS case directly, does not make sense. However,  the combinations similar to \eqref{relation virasoro BMS}, i.e.
\begin{equation}\label{combination}
T_1=\lim _{G/\ell\to 0}\frac{G}{\ell}(T_{++}+T_{--}),\qquad  T_2=\lim_{G/\ell\to0} (T_{++}-T_{--}),
\end{equation}
are finite in the flat limit. We define the energy-momentum tensor of flat-space , $\tilde T_{ij}$, by
\begin{equation}\label{def of T in flat}
T_1=(\tilde T_{++}+\tilde T_{--}),\qquad  T_2=(\tilde T_{++}-\tilde T_{--}),\qquad \tilde T_{+-}=0,
\end{equation}
where light-cone coordinates for flat spacetime,  $\tilde x^\pm$, are given by $\tilde x^\pm=u/G\pm\phi$. Using \eqref{def of T in flat} we see that 
\begin{equation}\label{def of EM1}
\tilde T_{uu}=\dfrac{{M}}{16\pi G^2},\qquad \tilde T_{u\phi}=\dfrac{{N}}{8\pi G^2},\qquad \tilde T_{\phi\phi}=\dfrac{{M}}{16\pi},
\end{equation}
are finite where $M$ and $N$ are given by \eqref{MN flat}. One can check that \eqref{def of EM1} is given by the following scaling from the energy-momentum tensor of AdS case: 
\begin{equation}
\tilde T_{uu}=\lim_{{G\over \ell}\to 0} {\ell\over G} T_{uu},\qquad \tilde T_{u\phi}=\lim_{{G\over \ell}\to 0} {\ell\over G} T_{u\phi},\qquad \tilde T_{\phi\phi}=\lim_{{G\over \ell}\to 0} {G\over \ell} T_{\phi\phi}.
\end{equation}

As mentioned in the section \eqref{BMS/GCA}, the holographic dual of asymptotically flat spacetimes is a field theory with Galilean symmetry. Thus the energy-momentum tensor which we define in this section must be consistent with what one expects to see for such a field theory. The first observation is that the combination \eqref{combination} is the same as what people define as the energy-momentum tensor of non-relativistic CFTs which are deduced by contracting a relativistic one( see for example \cite{Bagchi:2010vw}\footnote{In the definition of \cite{Bagchi:2010vw} the $\epsilon$ factor (reminsent of our $G/\ell$) appears in the subtraction of components of energy-momentum tensor. In fact in that paper GCA arises by contracting $x$-coordinate which is different from our case. }) . 

In fact the relation with an ultra-relativistic theory imposes that we change the conservation relation of energy-momentum tensor. To find the correct expression of conservation which our energy-momentum tensor must satisfy, let us indicate the coordinates of relativistic CFT on the cylinder with $(t,x)$ where $x$ is periodic and the radius of cylinder is absorbed in its periodicity. The light-cone coordinates are $x^\pm=t\pm x$. The field theory with Galilean symmetry is deducible from the relativistic one by contracting time as $t\to\epsilon t$ and taking the $\epsilon\to 0$ limit. According to \eqref{combination} and \eqref{def of T in flat}, for the components of energy momentum tensor of Galilean CFT, $\tilde T_{ij}$,  we have 
\begin{equation}
T_1=\tilde T_{++}+\tilde T_{--}=\lim_{\epsilon \to 0} \epsilon(T_{++}+T_{--}),\qquad T_2=\tilde T_{++}-\tilde T_ {--}=\lim_{\epsilon\to 0} (T_{++}-T_{--}),
\end{equation}
where $(+,-)$ in  $\tilde T_{ij}$ are defined by  contracted time. Now one can  check that the relativistic conservation of energy momentum tensor i.e. $\nabla^i T_{ij}=0$ reduces to 
\begin{equation}\label{non-relativistic conservation}
\partial_t T_1=0,\qquad \partial_x T_1-\partial_t T_2=0,
\end{equation}
where $t$ is the contracted time.
In the flat bulk side, $u$ corresponds to $t$ and $ G \phi$ corresponds to $x$. Thus using  \eqref{def of EM1} we can translate the ultra relativistic conservation relation \eqref{non-relativistic conservation} in terms of parameters of the bulk geometry as 
\begin{equation}\label{equation of MN}
\partial_u M=0,\qquad \partial_\phi M-2\partial _u N=0,
\end{equation}
which are satisfied by using values of $M$ and $N$ in \eqref{MN flat}. In fact \eqref{equation of MN} are exactly the equations which one finds by putting the ansatz \eqref{flat  in BMS} in the Einstein equations.

The proposal for dual boundary theory of asymptotically flat spacetimes as an ultra relativistic CFT, can be used for finding conserved charges. According to \eqref{BMS in ligh-cone}, the coordinates $(t,x)$ of parent CFT corresponds to $({G\over\ell}u, G\phi)$ in the bulk side.It is obvious from this identification  that ultra-relativistic contraction $t\to\epsilon t $ corresponds to the flat limit ${G\over \ell} \to 0$ in the bulk side. Thus if we denote by $(\tilde t,\tilde x)$ the coordinates of ultra-relativistic theory, they correspond to $(u,G\phi)$ in the bulk side. We deduce that the ultra-relativistic field theory must live on a flat spacetime $\partial M$ with line element  $ds^2=-du^2+G^2 d\phi^2$. Now we can use components of  our energy-momentum tensor \eqref{def of EM1} for finding  conserved charges of  symmetry generators \eqref{AKV flat} by making use of Brown and York definition \cite{Brown:1992br}
\begin{equation}
 Q_\xi=\int_\Sigma\, d\phi \sqrt{\sigma} v^\mu \xi^\nu \tilde T_{\mu\nu},
 \end{equation} 
where $\Sigma$ is the spacelike surface of $\partial M$ ($u=$ constant surface), $\sigma_{ab}$ is metric of $\Sigma$ i.e. $\sigma_{ab}dx^a dx^b= G^2 d\phi^2$ and $v^\mu$ is the unit timelike vector normal to $\Sigma$. Putting every things together we find
\begin{equation}
Q_{T,Y}={1\over 16\pi G}\int_0^{2\pi} d\phi \left(\theta T+2\chi Y\right),
\end{equation}
which is consistent with the result of \cite{Barnich:bmscft1}.

\section{Conclusion}
BMS/GCA is a correspondence between a theory of gravity in asymptotically flat spacetimes and a field theory with Galilean conformal symmetry. Finding a field theory with this symmetry may seem somewhat difficult at first but a route to this problem was found in  \cite{Bagchi:2012cy} where it was proposed that the field theory would be an ultra relativistic limit of a usual relativistic CFT. It was also shown how the flat limit in the bulk corresponds to this ultra relativistic contraction of the boundary CFT. 

In this paper we showed that this proposal can be used in defining the energy-momentum tensor of asymptotically flat spacetimes. Taking the flat limit from energy-momentum tensor of AdS case is not trivial in general. However, on the boundary side we know the relation between generators of conformal symmetry and GCA. This gives a good hint for defining  one point function of energy-momentum tensor of Galilean CFT. In this paper we  used such a relation and found reasonable expressions for the energy-momentum tensor of flat gravity. As expected, the corresponding energy-momentum tensor satisfies a conservation relation which is the ultra-relativistic version of relativistic equations. The fact that these conservation equations are similar to Einstein equations shows that our definition in the bulk side is consistent. We also used our energy-momentum tensor to calculate conserved charges of asymptotically flat spacetimes. The results are consistent with other methods. The point which we used in this calculation is that we implicitly assumed that the ultra-relativistic theory lives on a cylinder rather than null infinity. In other words, if we indicate metric of boundary by $ds^2=-dt^2+dx^2$  for the relativistic theory and contract time as $t=\epsilon t'$, the non-relativistic theory lives on a hyperspace with line element $ds^2=-dt'^2+dx^2$.  

We believe that this work is extendible to higher dimensions. For example one can find an energy-momentum tensor for producing mass and angular momentum of AdS-Kerr black holes.  The point is that one should write it in the BMS gauge and take its limit. Our lesson from three dimensional case is that in order to find the energy-momentum tensor in the context of flat holography one should start from BMS gauge and define the flat limit by constructing  a dimensionless parameter from $\ell$ and $G$ and scale all components of $T_{ij}$ written in the BMS-coordinate in such a way that $\ell$ disappears. In fact this way of constructing energy-momentum tensor has roots in the dual ultra-relativistic conformal field theory. We hope to address this problem in future work.   

\section*{Acknowledgement}
The authors  would like to specially thank A. E. Mosaffa  and A. Bagchi for their comments on the manuscript and revised version. We also would like to thank M. Alishahiha, D. Allahbakhshi and A. Vahedi for valuable discussions and useful comments during the course of this work.

\end{document}